\documentclass[aps,prl,preprint,showpacs]{revtex4-1}
\usepackage{graphicx}
\newcommand{\ket}[1]{{\left| {#1} \right\rangle}}

\begin{document}
\title{Ultrafast Gates for Single Atomic Qubits}
\date{\today}
\author{W. C. Campbell}
\email{wes3000@umd.edu}
\author{J. Mizrahi}
\author{Q. Quraishi}
\author{C. Senko}
\author{D. Hayes}
\author{D. Hucul}
\author{D. N. Matsukevich}
\author{P. Maunz}
\author{C. Monroe}
\affiliation{Joint Quantum Institute, University of Maryland
  Department of Physics and National Institute of Standards and
  Technology, College Park, Maryland 20742 USA}

\begin{abstract}
  We demonstrate single qubit operations on a trapped atom hyperfine
  qubit using a single ultrafast pulse from a mode-locked laser.  We
  shape the pulse from the laser and perform a $\pi$ rotation of the
  qubit in less than 50 ps with a population transfer exceeding 99\%
  and negligible effects from spontaneous emission or ac Stark
  shifts.  The gate time is significantly shorter than the period of
  atomic motion in the trap
  ($\Omega_{\mathrm{Rabi}}/\nu_{\mathrm{trap}}>10^4$),
  demonstrating that this interaction takes place deep within the
  strong excitation regime. 
\end{abstract}
\maketitle

Quantum information processing requires the ability to perform
operations in an amount of time shorter than the coherence time of the
qubit.  The ratio between coherence time and gate duration can be
increased by improving coherence times or developing faster gates.
One experimental extreme is found in atomic systems such as trapped
ions, which can have coherence times of many minutes
\cite{BollingerIEEETIM91, FiskIEEETUFFC97}.  However, the precise
spectral resolution needed to address narrow transitions and available
excitation power have typically limited gate times to around a
microsecond.  In many condensed matter systems, on the other hand,
single qubit operations must be performed on picosecond timescales
\cite{DuttPRB06, BerezovskySCIENCE08, FuNatPhys08} owing to the short
coherence time of the qubit ($T_2^{\ast}<10 \mbox{ ns}$ in
\cite{DuttPRB06, BerezovskySCIENCE08, FuNatPhys08}).  Here we realize
ultrafast operations in an atomic system with very slow decoherence
rates.

Aside from increasing the clock speed to decoherence rate ratio, there
are also specific applications that would benefit immediately from
fast single qubit gates, such as noise-reduction techniques
\cite{UhrigPRL07}, entanglement of material qubits with photon
time-bin qubits \cite{BarrettPRA05}, and probabilistic gates with
repetition rates limited by qubit rotations \cite{OlmschenkSCIENCE09}.
In particular, the ability to apply fast spin-dependent momentum kicks
to trapped ions in the strong excitation regime \cite{PoyatosPRA96} is
a critical ingredient for fast sideband cooling \cite{MachnesPRL10}
and fast entanglement of multiple atomic ion qubits
\cite{GarciaRipollPRL03, DuanPRL04}.  Such entangling gates can
be performed faster than a trap oscillation period, in contrast to
motional gates using spectroscopically resolved sidebands.

We previously reported the implementation of an optical frequency comb
to perform qubit operations \cite{HayesPRL10}.  In the low excitation
regime, each pulse alters the atomic state by a tiny amount and the
effect of the pulse train is manifested through careful coherent
accumulation of spectral density into narrow comb teeth, resulting in
gate times of order 100 $\mu$s.  In this Letter, we
realize the strong excitation regime ($\Omega_{\mathrm{Rabi}} \gg
\nu_{\mathrm{trap}}$) \cite{PoyatosPRA96} and demonstrate fast control
of an atomic qubit by driving stimulated Raman transitions with a
single pulse from a picosecond mode-locked laser.  In order to
implement a fast $\pi$-pulse, we perform simple pulse shaping in the
form of a beam splitter and delay line \cite{WeinerPQE95} to fully
rotate the qubit in less than $50 \mbox{ ps}$ with population transfer
exceeding 99\%.  Since the atomic qubit is well isolated from its
environment, the time required to perform such a fast gate is only a
small fraction ($< 10^{-8}$) of the measured coherence time of our
qubit.  By setting the delay to zero in a counter-propagating
geometry, the transition probability becomes sensitive to the atomic
motion, which is the first step toward implementing fast
cooling \cite{MachnesPRL10} and fast entangling gates between multiple
ions \cite{DuanPRL04, GarciaRipollPRL03}.

\begin{figure}
\includegraphics{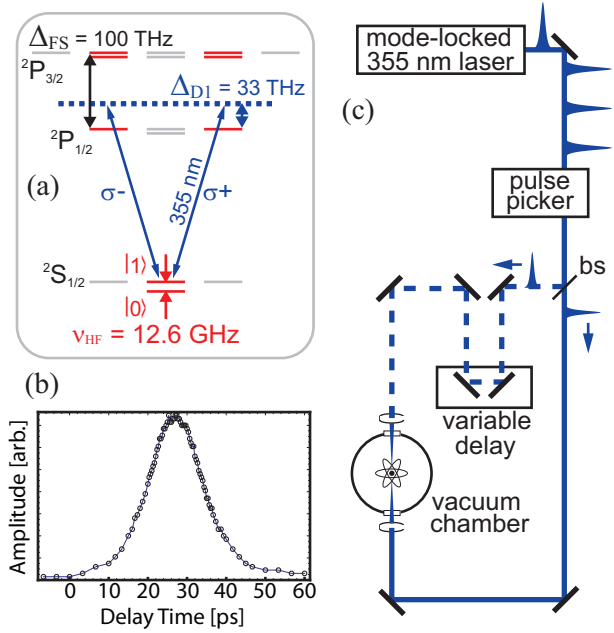}
\caption{Experimental schematic showing (a) the energy level diagram
  of $^{171}$Yb$^+$ with relevant states (not to scale), (b) the envelope of
  the electric field autocorrelation from the mode-locked laser and
  (c) the pulsed laser beam path with optional (dashed line)
  pulse-shaping beam path.\label{ExptDiagram}}
\end{figure}

Our experimental apparatus is shown in Fig. \ref{ExptDiagram}(c).  A
$^{171}$Yb$^+$ ion is trapped in a linear radio frequency Paul trap.
The $m_{F}=0$ ``clock states'' of the two $^2\mathsf{S}_{1/2}$
hyperfine levels act as the basis states of our qubit separated in
frequency by $\nu_{\mathrm{HF}}=12.642815 \mbox{ GHz}$.  Doppler
cooling, state preparation, and state detection are accomplished on
the $^2\mathsf{P}_{1/2} \leftrightarrow {^2}\mathsf{S}_{1/2}$ ``D1
line'' at 370 nm as described in Ref. \cite{OlmschenkPRA07}.  The
mode-locked laser for Raman transitions is a frequency-tripled yttrium
vanadate laser operating at a repetition rate of
$f_{\mathrm{rep.}}=121$ MHz with an average power of 4 W at 355 nm
\footnote{Coherent Paladin Compact 355-4000}.  Single pulses are
extracted by a Pockels cell pulse picker \footnote{Fast Pulse
  Technology model 5046SC} and focused onto the ion.

Fig. \ref{ExptDiagram}(a) shows the relevant energy levels of
$^{171}$Yb$^+$.  The $^2\mathsf{S}_{1/2}$ qubit states
$\ket{0}$ and $\ket{1}$ are coupled through excited $^2\mathsf{P}$
states via far-detuned light that is polarized to drive either
$\sigma^+$ or $\sigma^-$ transitions, as Raman transitions from $\pi$
light are forbidden by selection rules.  The $\sigma^{\pm}$ Raman
transition (two-photon) Rabi frequency is
\cite{WinelandPTRSLA03, OzeriPRA07}
\begin{equation}
\Omega_{0,1} = \pm \frac{g^2}{6}\left( \frac{1}{\Delta_{\mathrm{D1}}} +
                 \frac{1}{\Delta_{\mathrm{FS}} - \Delta_{\mathrm{D1}}}\right)  ,
\label{OmegaRaman}
\end{equation}
where $g$ is the resonant one-photon Rabi frequency of the
$^2\mathsf{P}_{3/2}(F=2, m_F=2) \leftrightarrow
{^2}\mathsf{S}_{1/2}(F=1, m_F=1)$ ``D2 line'' cycling transition,
$\Delta_{\mathrm{D1}}$ is the detuning of the light above the
$^2\mathsf{P}_{1/2}$ state,
$\Delta_{\mathrm{FS}}-\Delta_{\mathrm{D1}}$ is the detuning of the
light below the $^2\mathsf{P}_{3/2}$ state, and $\Delta_{\mathrm{FS}}
\approx 100$ THz is the fine structure splitting.  The ac Stark
shifts of the qubit states from $\sigma^{\pm}$ light are
\cite{WinelandPTRSLA03, OzeriPRA07}
\begin{eqnarray} 
\delta_0&=& \frac{g^2}{12}\left( \frac{1}{\Delta_{\mathrm{D1}}} -
                                          \frac{2}{\Delta_{\mathrm{FS}} - \Delta_{\mathrm{D1}}}\right) \\
\delta_1&=& \frac{g^2}{12}\left( \frac{1}{\Delta_{\mathrm{D1}}+\nu_{\mathrm{HF}}} -
                                          \frac{2}{\Delta_{\mathrm{FS}} - \Delta_{\mathrm{D1}}-\nu_{\mathrm{HF}}}\right) ,
\label{ACStark}
\end{eqnarray}
where we have neglected the hyperfine splitting of the
$^2\mathsf{P}_{3/2}$ state.

Eqs. \ref{OmegaRaman}-\ref{ACStark} show that if the Raman laser is
tuned between the D1 and D2 lines
($0<\Delta_{\mathrm{D1}}<\Delta_{\mathrm{FS}}$) the stimulated Raman
transition amplitudes due to the $^2\mathsf{P}_{1/2}$ and
$^2\mathsf{P}_{3/2}$ couplings add constructively while the ac Stark
shift contributions interfere destructively.  The Stark shifts of the
qubit states each cross zero near an optimal wavelength of
$\Delta_{\mathrm{D1}}^{\mathrm{opt.}} =\Delta_{\mathrm{FS}}/3$, which
for Yb$^+$ corresponds to a Raman laser wavelength of
$\lambda_{\mathrm{opt.}} = 355 \mbox{ nm}$.  The differential ac Stark
shift of the qubit states does not exactly cancel, but has a local
minimum near $\lambda_{\mathrm{opt.}}$, reaching a value of $\delta_0
- \delta_1 \approx \Omega_{0,1} \times 3\nu_{\mathrm{HF}}/2\Delta_{\mathrm{FS}}$,
corresponding to $2 \times 10^{-4} \,\,\Omega_{0,1}$ for Yb$^+$.  We
measure a differential Stark shift of $1.1(5) \times 10^{-4}
\,\,\Omega_{0,1}$ through microwave Ramsey spectroscopy with
linearly polarized 355 nm light.

The spontaneous emission rate can be estimated by calculating the
excited state populations during the Raman transition to give
\cite{WinelandPTRSLA03, OzeriPRA07}
\begin{equation}
\Gamma_{\mathrm{spon}} = \gamma \frac{g^2}{6}\left( \frac{1}{\Delta_{\mathrm{D1}}^2}+ \frac{2}{(\Delta_{\mathrm{FS}} - \Delta_{\mathrm{D1}})^2} \right),\label{GammaSpon}
\end{equation}
where $\gamma \sim 2 \pi \times 20$ MHz is the spontaneous emission
rate from the $^2\mathsf{P}$ states \footnote{For
  Eq. \ref{OmegaRaman}-\ref{GammaSpon} we have ignored the small
  contributions from the $^3[3/2]^{\mathrm{o}}_{3/2}$ state transition
  at 348 nm caused by the $^2\mathsf{P}_{3/2}$ admixture in that
  state.  We have also neglected the difference in transition dipole
  moment between the D1 and D2 lines.  Both effects contribute less
  than 15\% corrections to the two-photon Rabi frequency, differential
  Stark shift, and spontaneous emission rate at 355 nm.}.  The
probability of a spontaneous emission event during a $\pi$-pulse near
$\lambda_{\mathrm{opt.}}$ can be estimated as
$\mathcal{P}_{\mathrm{spon,}\pi} \approx 3 \gamma/ 2
\Delta_{\mathrm{FS}}$.  The local minima in $\Gamma_{\mathrm{spon}}$
and $\mathcal{P}_{\mathrm{spon,}\pi}$ lie close to
$\lambda_{\mathrm{opt.}}$ for Yb$^+$ at 349 nm and 352 nm,
respectively.  At 355 nm we estimate
$\mathcal{P}_{\mathrm{spon,}\pi}<10^{-5}$.

In order to operate in this low Stark shift, low spontaneous emission
regime, sufficient optical power at $\lambda_{\mathrm{opt.}}$ must be
provided to make $\Omega_{0,1}$ appreciable.  For Yb$^+$, frequency
tripled mode-locked Nd:YAG and vanadate lasers are available that
provide many Watts at a center wavelength of 355 nm \footnote{Despite
  being slightly outside the fine-structure interval, a laser at 532
  nm ($= 1064 \mbox{ nm}/2$) has a two-photon Rabi frequency for
  Ba$^+$ that is only slightly lower than Yb$^+$ at 355 nm with
  comparable pulse energy.}.

The two-photon Rabi frequency in Eq. \ref{OmegaRaman} is
time-dependent due to the shape of the pulse.  As long as the pulse
bandwidth is small compared to the single photon detunings $\Delta$,
the adiabatic elimination of the excited states remains valid and we
can treat the qubit as a two-level system with a time-dependent
coupling:
\begin{equation}
H_{\mathrm{eff}}/h=-\frac{\nu_{\mathrm{HF}}}{2}\hat{\sigma}_z -
\frac{\Omega_{0,1}(t)}{2} \hat{\sigma}_x\label{Heff}.
\end{equation}
Numerical solutions to the Schr\"{o}dinger equation with this
Hamiltonian can be obtained for a general time-dependent $\Omega(t)$.
However, there is an analytic solution due to Rosen and Zener for a
coupling with a hyperbolic secant time-dependence \cite{RosenPR32},
which is the electric field envelope expected from picosecond
mode-locked pulses \cite{Siegman}.  Fig. \ref{ExptDiagram}(b) shows
the envelope of an electric field autocorrelation of a single pulse as
measured with a scanning Mach-Zender interferometer and a fast
photodiode, which is consistent with the autocorrelation of a
hyperbolic secant with $T_{\mathrm{pulse}}=14.8 \mbox{ ps}$ having a
linear pulse chirp of $8 \times 10^{-3} \mbox{ ps}^{-2}$.
Eq. \ref{OmegaRaman}, however, shows that the two-photon Rabi
frequency is proportional to the square of the electric field
envelope.  Nonetheless, $\mathrm{sech}^2$ and $\mathrm{sech}$ are
sufficiently similar that numerical solutions to the Schr\"{o}dinger
equation with $\Omega(t) \propto \mathrm{sech}^2[2 t/T]$ match the
analytic solutions with $\Omega(t) \propto \mathrm{sech} [\pi t/T]$ to
within a few percent for all simulations shown here.  The Rosen-Zener
transition probability from an initial state $\ket{0}$ to $\ket{1}$ is
\cite{RosenPR32}
\begin{equation}
\mathcal{P}_{0 \rightarrow 1} =
\sin^2(\pi u)\,\,\mathrm{sech}^2[\pi \nu_{\mathrm{HF}}T_{\mathrm{pulse}}]\label{RosenZenerProbability}
\end{equation}
for a two-photon Rabi frequency of $\Omega_{1,0}(t) =
\frac{u}{T_{\mathrm{pulse}}} \mathrm{sech}[\frac{\pi
  t}{T_{\mathrm{pulse}}}]$ where $u$ is the pulse area.
Eq. \ref{RosenZenerProbability} indicates that in order to create a
high-fidelity $\pi$-pulse, the pulse duration must be many times
shorter than the hyperfine period.

\begin{figure}
\includegraphics{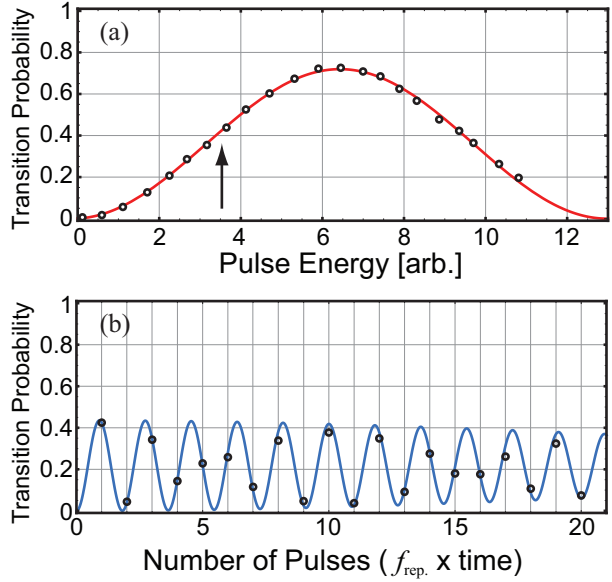}
\caption{(a) Qubit transition probability from a single pulse
  vs. pulse energy.  The solid curve is a fit to
  Eq. \ref{RosenZenerProbability} and the peak corresponds to a pulse
  energy of $12 \pm 2$ nJ.  (b) Qubit transition probability vs. the
  number of identical ($\approx 8 \mbox{ nJ}$) pulses at the laser
  repetition rate.  The solid curve is a fit to the Rosen-Zener
  solution with a Gaussian contrast decay.  The single pulse energy
  for (b) is shown with an arrow in (a).\label{SinglePulseBrightness}}
\end{figure}

Fig. \ref{SinglePulseBrightness}(a) shows the measured transition
probability as a function of pulse energy (proportional to $u$ in
Eq. \ref{RosenZenerProbability} and monitored with a fast photodiode).
Both plots in Fig. \ref{SinglePulseBrightness} were taken with a
single beam path, as represented by the solid line in
Fig. \ref{ExptDiagram}(c).  The maximum transition probability of 72\%
corresponds (see Eq. \ref{RosenZenerProbability}) to a pulse duration
of $T_{\mathrm{pulse}}= 14.8 \mbox{ ps}$.

We also studied the dynamics of a train of identical pulses (with
pulse energy denoted by the arrow in
Fig. \ref{SinglePulseBrightness}(a)) at the laser repetition rate,
shown in Fig. \ref{SinglePulseBrightness}(b).  The solid curve is an
analytic solution including free evolution with a fitted Gaussian
decay in contrast.  The fitted pulse duration is
$T_{\mathrm{pulse}}=14 \mbox{ ps}$, in good agreement with the fitted
pulse width from Fig. \ref{SinglePulseBrightness}(a).

It is clear from Eq. \ref{RosenZenerProbability} and
Fig. \ref{SinglePulseBrightness}(a) that for $T_{\mathrm{pulse}}=14.8
\mbox{ ps}$, even with unlimited pulse energy the population transfer
probability cannot exceed 72\% without changing the pulse
shape.  In order to execute a full $\pi$-pulse, we introduced a simple
pulse-shaping element into the beam path in the form of a beamsplitter
and a delay line \cite{WeinerPQE95}, resulting in two
counter-propagating pulses as shown in Fig. \ref{ExptDiagram}(c).  By
setting the energy of each half of the pulse to transfer 50\% of the
population, we are able to perform a fast Ramsey experiment with a
variable free evolution time between two $\pi/2$ Raman pulses.  A
$\pi$-pulse then corresponds to the top of a Ramsey fringe.

\begin{figure}
\includegraphics{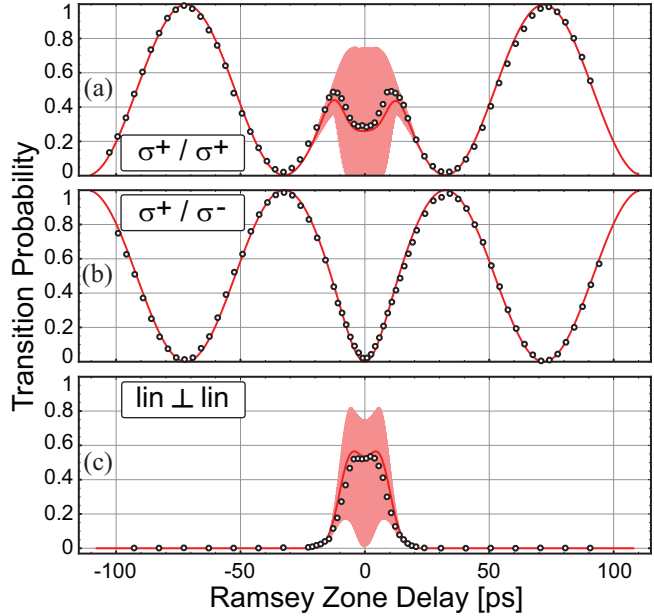}
\caption{Ramsey fringes for three different polarization orientations
  for each half of the shaped pulse obtained by fitting ensemble
  histograms of photon counts.  The polarization configurations are
  (a) pure $\sigma^+$ light, (b) one is $\sigma^+$ and the other
  $\sigma^-$, and (c) lin $\perp$ lin.  The overlap region near zero
  delay results in an optical standing wave with a period of 177 nm.
  The solid curves of (a) and (c) are calculations based on
  independent measurements of the optical pulse chirp and envelope
  with no free fit parameters.  The bold curve also incorporates a
  thermal average of the intial ion position at the Doppler cooling
  limit of $\bar{n}=40$, washing out the fine standing wave
  fringes.\label{RamseyFringes1}}
\end{figure}

The results of the pulse shape delay scan are shown in
Fig. \ref{RamseyFringes1}(a)-(c).  We show results from experiments
with three different polarization orientations for the two
$\pi/2$-pulses.  In (a), both beams drive only $\sigma^+$ transitions.
The Ramsey fringes are first maximized for a delay of 72 ps,
corresponding to a net $\pi$-pulse after 80 ps.  In the frequency
domain, the shaped pulse spectrum in this case is a crude frequency
comb with sinusoidal teeth separated by 14 GHz.  The reason this
separation is not exactly $12.6 \mbox{ GHz} = \nu_{\mathrm{HF}}$ is
the additional $\hat{\sigma}_z$ rotation introduced by the Rosen-Zener
solution dynamics.  In the limit of infinitesimally short pulses, the
comb spacing would converge to $\nu_{\mathrm{HF}}$.
Fig. \ref{RamseyFringes1}(b) shows the same experiment with one beam
driving $\sigma^+$ transitions and the other driving $\sigma^-$.  The
two central peaks each represent a $\sim 40$ ps net $\pi$-pulse, which
is shorter than the 80 ps pure $\sigma^+$ $\pi$-pulse due to the fact
that the Rabi frequency for $\sigma^+$ and $\sigma^-$ transitions have
opposite sign and therefore rotate the Bloch vector about different
axes.  Fig. \ref{RamseyFringes1}(c) shows the same experiment with
orthogonal linear polarizations (``lin $\perp$ lin") where neither
beam drives Raman transitions by itself.

In order to quantify the population transfer probability, we compared
detection histograms \cite{OlmschenkPRA07} from a fast Raman
$\pi$-pulse to a $\pi$-pulse applied with microwaves and measured a
population transfer probability of 99.3\%.  We also investigated the
population transfer efficiency of the $\pi$-pulses by repeating the
gate multiple times at the laser repetition rate and monitoring the
transfer probability while increasing the number of $\pi$-pulses.
Since the laser repetition period is very close to a half integer
number of hyperfine evolution periods
($\nu_{\mathrm{HF}}/f_{\mathrm{rep.}}\approx 104.5$), there is some
natural spin-echo-type error cancellation for multiple pulses for our
system.  For the $\sigma^+$/$\sigma^+$ configuration, we measure a
population transfer probability contrast of 91\% after 25 $\pi$-pulses
while for $\sigma^+$/$\sigma^-$ we measure 77\%.

In the overlap region (near zero delay time) in
Fig. \ref{RamseyFringes1}, the two counterpropagating $\pi/2$-pulses
begin to overlap in time at the position of the atom and form an
optical standing wave.  For the $\sigma^+$/$\sigma^+$ and lin $\perp$
lin configurations, the two-photon Rabi frequency has 177 nm period
spatial interference fringes.  In the overlap regions of
Fig. \ref{RamseyFringes1}(a) and (c), Raman transitions are sensitive
to the motion of the ion since momentum kicks of $2\hbar k$ are being
transferred from the optical field.  The pulse's spectrum, however, is
much wider than the trap frequency (100 GHz compared to 500 kHz), so
the pulse simultaneously drives many motional sidebands (including the
carrier transition).  For the $\sigma^+$/$\sigma^-$ configuration
(Fig. \ref{RamseyFringes1}(b)), the overlap region contains linear
polarization and does not drive Raman transitions.

The thin solid curves in Fig. \ref{RamseyFringes1} show the results of
a numerical solution of the Schr\"{o}dinger equation vs. delay, which
oscillate at optical-frequency delays of less than 1 fs in the overlap
region.  In this region, the ion experiences momentum kicks and phase
shifts as spin states spread out in motional phase space, leaving the
interferometer open.  Since the final spin phase of different motional
states depends sensitively on the ion's intial position, thermal
averaging tends to wash out the fast spatial variation of the
transition probability. The thick curves show a thermal average over
the ion's initial state, assumed to be a thermal state at the Doppler
cooling limit of $\bar{n}=40$. The thin and thick curves in
Fig. \ref{RamseyFringes1} are not fit to the data as these numerical
solutions are fully constrained by the pulse duration and chirp
measurements shown in Fig. \ref{ExptDiagram}(b) and
\ref{SinglePulseBrightness}(a).

Implementation of the fast cooling \cite{MachnesPRL10} and fast
entangling gates \cite{GarciaRipollPRL03, DuanPRL04} will require
repeated spin-dependent momentum kicks generated through ultrafast
interferometry in the strong excitation regime \cite{PoyatosPRA96}.
In this case, the interferometer will be closed and therefore will not
be sensitive to optical wavelength interference such as the overlap
region in Fig. \ref{RamseyFringes1}(c).

We have demonstrated ultrafast single-qubit gates with a mode-locked
laser pulse using an atomic qubit. For a single trapped ion, the
free-evolution of the qubit can be used to perform $\hat{\sigma}_z$
rotations, and delaying the pulse arrival time will allow a rotation
about an arbitrary axis in the $x$-$y$ plane of the Bloch sphere.  As
such, the fundamental limit on the gate speed is the hyperfine period
(analogous to the Larmor precession time), which would yield a gate
time of $\approx 100$ ps.  Previous results with these same qubit
levels in $^{171}$Yb$^+$ have demonstrated coherence times in excess
of 1000 s \cite{FiskIEEETUFFC97}, so this single-qubit gate can be
performed in a vanishingly small fraction ($< 10^{-13}$) of the
coherence time.

\begin{acknowledgments}
  We acknowledge helpful discussions with Michael Biercuk, Ming-Shien
  Chang, Kihwan Kim, and Steven Olmschenk. This work is supported by
  the ARO with funds from the DARPA Optical Lattice Emulator (OLE)
  Program, IARPA under ARO contract, the NSF Physics at the
  Information Frontier Program, the IC Postdoctoral Program, and the
  NSF Physics Frontier Center at JQI.
\end{acknowledgments}

\bibliography{Ultrafast_Strong_Pulse_Bib}

\begin{thebibliography}{23}%
\makeatletter
\providecommand \@ifxundefined [1]{%
 \@ifx{#1\undefined}
}%
\providecommand \@ifnum [1]{%
 \ifnum #1\expandafter \@firstoftwo
 \else \expandafter \@secondoftwo
 \fi
}%
\providecommand \@ifx [1]{%
 \ifx #1\expandafter \@firstoftwo
 \else \expandafter \@secondoftwo
 \fi
}%
\providecommand \natexlab [1]{#1}%
\providecommand \enquote  [1]{``#1''}%
\providecommand \bibnamefont  [1]{#1}%
\providecommand \bibfnamefont [1]{#1}%
\providecommand \citenamefont [1]{#1}%
\providecommand \href@noop [0]{\@secondoftwo}%
\providecommand \href [0]{\begingroup \@sanitize@url \@href}%
\providecommand \@href[1]{\@@startlink{#1}\@@href}%
\providecommand \@@href[1]{\endgroup#1\@@endlink}%
\providecommand \@sanitize@url [0]{\catcode `\\12\catcode `\$12\catcode
  `\&12\catcode `\#12\catcode `\^12\catcode `\_12\catcode `\%12\relax}%
\providecommand \@@startlink[1]{}%
\providecommand \@@endlink[0]{}%
\providecommand \url  [0]{\begingroup\@sanitize@url \@url }%
\providecommand \@url [1]{\endgroup\@href {#1}{\urlprefix }}%
\providecommand \urlprefix  [0]{URL }%
\providecommand \Eprint [0]{\href }%
\@ifxundefined \urlstyle {%
  \providecommand \doi  [0]{\begingroup \@sanitize@url \@doi}%
  \providecommand \@doi [1]{\endgroup \@@startlink {\doibase
  #1}doi:\discretionary {}{}{}#1\@@endlink }%
}{%
  \providecommand \doi  [0]{doi:\discretionary{}{}{}\begingroup
  \urlstyle{rm}\Url }%
}%
\providecommand \doibase [0]{http://dx.doi.org/}%
\providecommand \Doi [0]{\begingroup \@sanitize@url \@Doi }%
\providecommand \@Doi  [1]{\endgroup\@@startlink{\doibase#1}\@@Doi}%
\providecommand \@@Doi [1]{#1\@@endlink}%
\providecommand \selectlanguage [0]{\@gobble}%
\providecommand \bibinfo  [0]{\@secondoftwo}%
\providecommand \bibfield  [0]{\@secondoftwo}%
\providecommand \translation [1]{[#1]}%
\providecommand \BibitemOpen [0]{}%
\providecommand \bibitemStop [0]{}%
\providecommand \bibitemNoStop [0]{.\EOS\space}%
\providecommand \EOS [0]{\spacefactor3000\relax}%
\providecommand \BibitemShut  [1]{\csname bibitem#1\endcsname}%
\bibitem [{\citenamefont {Bollinger}\ \emph {et~al.}(1991)\citenamefont
  {Bollinger}, \citenamefont {Heinzen}, \citenamefont {Itano}, \citenamefont
  {Gilbert},\ and\ \citenamefont {Wineland}}]{BollingerIEEETIM91}%
  \BibitemOpen
  \bibfield  {author} {\bibinfo {author} {\bibfnamefont {J.~J.}\ \bibnamefont
  {Bollinger}}, \bibinfo {author} {\bibfnamefont {D.~J.}\ \bibnamefont
  {Heinzen}}, \bibinfo {author} {\bibfnamefont {W.~M.}\ \bibnamefont {Itano}},
  \bibinfo {author} {\bibfnamefont {S.~L.}\ \bibnamefont {Gilbert}}, \ and\
  \bibinfo {author} {\bibfnamefont {D.~J.}\ \bibnamefont {Wineland}},\
  }\href@noop {} {\bibfield  {journal} {\bibinfo  {journal} {IEEE Trans. Instr.
  Meas.},\ }\textbf {\bibinfo {volume} {40}},\ \bibinfo {pages} {126} (\bibinfo
  {year} {1991})}\BibitemShut {NoStop}%
\bibitem [{\citenamefont {Fisk}\ \emph {et~al.}(1997)\citenamefont {Fisk},
  \citenamefont {Sellars}, \citenamefont {Lawn},\ and\ \citenamefont
  {Coles}}]{FiskIEEETUFFC97}%
  \BibitemOpen
  \bibfield  {author} {\bibinfo {author} {\bibfnamefont {P.~T.~H.}\
  \bibnamefont {Fisk}}, \bibinfo {author} {\bibfnamefont {M.~J.}\ \bibnamefont
  {Sellars}}, \bibinfo {author} {\bibfnamefont {M.~A.}\ \bibnamefont {Lawn}}, \
  and\ \bibinfo {author} {\bibfnamefont {C.}~\bibnamefont {Coles}},\
  }\href@noop {} {\bibfield  {journal} {\bibinfo  {journal} {IEEE Trans.
  Ultrason. Ferroelectr. Freq. Control},\ }\textbf {\bibinfo {volume} {44}},\
  \bibinfo {pages} {344} (\bibinfo {year} {1997})}\BibitemShut {NoStop}%
\bibitem [{\citenamefont {Dutt}\ \emph {et~al.}(2006)\citenamefont {Dutt},
  \citenamefont {Cheng}, \citenamefont {Wu}, \citenamefont {Xu}, \citenamefont
  {Steel}, \citenamefont {Bracker}, \citenamefont {Gammon}, \citenamefont
  {Economou}, \citenamefont {Liu},\ and\ \citenamefont {Sham}}]{DuttPRB06}%
  \BibitemOpen
  \bibfield  {author} {\bibinfo {author} {\bibfnamefont {M.~V.~G.}\
  \bibnamefont {Dutt}}, \bibinfo {author} {\bibfnamefont {J.}~\bibnamefont
  {Cheng}}, \bibinfo {author} {\bibfnamefont {Y.}~\bibnamefont {Wu}}, \bibinfo
  {author} {\bibfnamefont {X.}~\bibnamefont {Xu}}, \bibinfo {author}
  {\bibfnamefont {D.~G.}\ \bibnamefont {Steel}}, \bibinfo {author}
  {\bibfnamefont {A.~S.}\ \bibnamefont {Bracker}}, \bibinfo {author}
  {\bibfnamefont {D.}~\bibnamefont {Gammon}}, \bibinfo {author} {\bibfnamefont
  {S.~E.}\ \bibnamefont {Economou}}, \bibinfo {author} {\bibfnamefont {R.-B.}\
  \bibnamefont {Liu}}, \ and\ \bibinfo {author} {\bibfnamefont {L.~J.}\
  \bibnamefont {Sham}},\ }\href@noop {} {\bibfield  {journal} {\bibinfo
  {journal} {Physical Review B},\ }\textbf {\bibinfo {volume} {74}},\ \bibinfo
  {pages} {125306} (\bibinfo {year} {2006})}\BibitemShut {NoStop}%
\bibitem [{\citenamefont {Berezovsky}\ \emph {et~al.}(2008)\citenamefont
  {Berezovsky}, \citenamefont {Mikkelsen}, \citenamefont {Stoltz},
  \citenamefont {Coldren},\ and\ \citenamefont
  {Awschalom}}]{BerezovskySCIENCE08}%
  \BibitemOpen
  \bibfield  {author} {\bibinfo {author} {\bibfnamefont {J.}~\bibnamefont
  {Berezovsky}}, \bibinfo {author} {\bibfnamefont {M.~H.}\ \bibnamefont
  {Mikkelsen}}, \bibinfo {author} {\bibfnamefont {N.~G.}\ \bibnamefont
  {Stoltz}}, \bibinfo {author} {\bibfnamefont {L.~A.}\ \bibnamefont {Coldren}},
  \ and\ \bibinfo {author} {\bibfnamefont {D.~D.}\ \bibnamefont {Awschalom}},\
  }\href@noop {} {\bibfield  {journal} {\bibinfo  {journal} {Science},\
  }\textbf {\bibinfo {volume} {320}},\ \bibinfo {pages} {349} (\bibinfo {year}
  {2008})}\BibitemShut {NoStop}%
\bibitem [{\citenamefont {Fu}\ \emph {et~al.}(2008)\citenamefont {Fu},
  \citenamefont {Clark}, \citenamefont {Santori}, \citenamefont {Stanley},
  \citenamefont {Holland},\ and\ \citenamefont {Yamamoto}}]{FuNatPhys08}%
  \BibitemOpen
  \bibfield  {author} {\bibinfo {author} {\bibfnamefont {K.-M.~C.}\
  \bibnamefont {Fu}}, \bibinfo {author} {\bibfnamefont {S.~M.}\ \bibnamefont
  {Clark}}, \bibinfo {author} {\bibfnamefont {C.}~\bibnamefont {Santori}},
  \bibinfo {author} {\bibfnamefont {C.~R.}\ \bibnamefont {Stanley}}, \bibinfo
  {author} {\bibfnamefont {M.~C.}\ \bibnamefont {Holland}}, \ and\ \bibinfo
  {author} {\bibfnamefont {Y.}~\bibnamefont {Yamamoto}},\ }\href@noop {}
  {\bibfield  {journal} {\bibinfo  {journal} {Nature Physics},\ }\textbf
  {\bibinfo {volume} {4}},\ \bibinfo {pages} {780} (\bibinfo {year}
  {2008})}\BibitemShut {NoStop}%
\bibitem [{\citenamefont {Uhrig}(2007)}]{UhrigPRL07}%
  \BibitemOpen
  \bibfield  {author} {\bibinfo {author} {\bibfnamefont {G.~S.}\ \bibnamefont
  {Uhrig}},\ }\href@noop {} {\bibfield  {journal} {\bibinfo  {journal} {Phys.
  Rev. Lett.},\ }\textbf {\bibinfo {volume} {98}},\ \bibinfo {pages} {100504}
  (\bibinfo {year} {2007})}\BibitemShut {NoStop}%
\bibitem [{\citenamefont {Barrett}\ and\ \citenamefont
  {Kok}(2005)}]{BarrettPRA05}%
  \BibitemOpen
  \bibfield  {author} {\bibinfo {author} {\bibfnamefont {S.~D.}\ \bibnamefont
  {Barrett}}\ and\ \bibinfo {author} {\bibfnamefont {P.}~\bibnamefont {Kok}},\
  }\href@noop {} {\bibfield  {journal} {\bibinfo  {journal} {Phys. Rev. A},\
  }\textbf {\bibinfo {volume} {71}},\ \bibinfo {pages} {060310(R)} (\bibinfo
  {year} {2005})}\BibitemShut {NoStop}%
\bibitem [{\citenamefont {Olmschenk}\ \emph {et~al.}(2009)\citenamefont
  {Olmschenk}, \citenamefont {Matsukevich}, \citenamefont {Maunz},
  \citenamefont {Hayes}, \citenamefont {Duan},\ and\ \citenamefont
  {Monroe}}]{OlmschenkSCIENCE09}%
  \BibitemOpen
  \bibfield  {author} {\bibinfo {author} {\bibfnamefont {S.}~\bibnamefont
  {Olmschenk}}, \bibinfo {author} {\bibfnamefont {D.~N.}\ \bibnamefont
  {Matsukevich}}, \bibinfo {author} {\bibfnamefont {P.}~\bibnamefont {Maunz}},
  \bibinfo {author} {\bibfnamefont {D.}~\bibnamefont {Hayes}}, \bibinfo
  {author} {\bibfnamefont {L.~M.}\ \bibnamefont {Duan}}, \ and\ \bibinfo
  {author} {\bibfnamefont {C.}~\bibnamefont {Monroe}},\ }\href@noop {}
  {\bibfield  {journal} {\bibinfo  {journal} {Science},\ }\textbf {\bibinfo
  {volume} {323}},\ \bibinfo {pages} {486} (\bibinfo {year}
  {2009})}\BibitemShut {NoStop}%
\bibitem [{\citenamefont {Poyatos}\ \emph {et~al.}(1996)\citenamefont
  {Poyatos}, \citenamefont {Cirac}, \citenamefont {Blatt},\ and\ \citenamefont
  {Zoller}}]{PoyatosPRA96}%
  \BibitemOpen
  \bibfield  {author} {\bibinfo {author} {\bibfnamefont {J.~F.}\ \bibnamefont
  {Poyatos}}, \bibinfo {author} {\bibfnamefont {J.~I.}\ \bibnamefont {Cirac}},
  \bibinfo {author} {\bibfnamefont {R.}~\bibnamefont {Blatt}}, \ and\ \bibinfo
  {author} {\bibfnamefont {P.}~\bibnamefont {Zoller}},\ }\href@noop {}
  {\bibfield  {journal} {\bibinfo  {journal} {Phys. Rev. A},\ }\textbf
  {\bibinfo {volume} {54}},\ \bibinfo {pages} {1532} (\bibinfo {year}
  {1996})}\BibitemShut {NoStop}%
\bibitem [{\citenamefont {Machnes}\ \emph {et~al.}(2010)\citenamefont
  {Machnes}, \citenamefont {Plenio}, \citenamefont {Reznik}, \citenamefont
  {Steane},\ and\ \citenamefont {Retzker}}]{MachnesPRL10}%
  \BibitemOpen
  \bibfield  {author} {\bibinfo {author} {\bibfnamefont {S.}~\bibnamefont
  {Machnes}}, \bibinfo {author} {\bibfnamefont {M.~B.}\ \bibnamefont {Plenio}},
  \bibinfo {author} {\bibfnamefont {B.}~\bibnamefont {Reznik}}, \bibinfo
  {author} {\bibfnamefont {A.~M.}\ \bibnamefont {Steane}}, \ and\ \bibinfo
  {author} {\bibfnamefont {A.}~\bibnamefont {Retzker}},\ }\href@noop {}
  {\bibfield  {journal} {\bibinfo  {journal} {Phys. Rev. Lett.},\ }\textbf
  {\bibinfo {volume} {104}},\ \bibinfo {pages} {183001} (\bibinfo {year}
  {2010})}\BibitemShut {NoStop}%
\bibitem [{\citenamefont {Garc\'{i}a-Ripoll}\ \emph {et~al.}(2003)\citenamefont
  {Garc\'{i}a-Ripoll}, \citenamefont {Zoller},\ and\ \citenamefont
  {Cirac}}]{GarciaRipollPRL03}%
  \BibitemOpen
  \bibfield  {author} {\bibinfo {author} {\bibfnamefont {J.~J.}\ \bibnamefont
  {Garc\'{i}a-Ripoll}}, \bibinfo {author} {\bibfnamefont {P.}~\bibnamefont
  {Zoller}}, \ and\ \bibinfo {author} {\bibfnamefont {J.~I.}\ \bibnamefont
  {Cirac}},\ }\href@noop {} {\bibfield  {journal} {\bibinfo  {journal} {Phys.
  Rev. Lett.},\ }\textbf {\bibinfo {volume} {91}},\ \bibinfo {pages} {157901}
  (\bibinfo {year} {2003})}\BibitemShut {NoStop}%
\bibitem [{\citenamefont {Duan}(2004)}]{DuanPRL04}%
  \BibitemOpen
  \bibfield  {author} {\bibinfo {author} {\bibfnamefont {L.-M.}\ \bibnamefont
  {Duan}},\ }\href@noop {} {\bibfield  {journal} {\bibinfo  {journal} {Phys.
  Rev. Lett.},\ }\textbf {\bibinfo {volume} {93}},\ \bibinfo {pages} {100502}
  (\bibinfo {year} {2004})}\BibitemShut {NoStop}%
\bibitem [{\citenamefont {Hayes}\ \emph {et~al.}(2010)\citenamefont {Hayes},
  \citenamefont {Matsukevich}, \citenamefont {Maunz}, \citenamefont {Hucul},
  \citenamefont {Quraishi}, \citenamefont {Olmschenk}, \citenamefont
  {Campbell}, \citenamefont {Mizrahi}, \citenamefont {Senko},\ and\
  \citenamefont {Monroe}}]{HayesPRL10}%
  \BibitemOpen
  \bibfield  {author} {\bibinfo {author} {\bibfnamefont {D.}~\bibnamefont
  {Hayes}}, \bibinfo {author} {\bibfnamefont {D.~N.}\ \bibnamefont
  {Matsukevich}}, \bibinfo {author} {\bibfnamefont {P.}~\bibnamefont {Maunz}},
  \bibinfo {author} {\bibfnamefont {D.}~\bibnamefont {Hucul}}, \bibinfo
  {author} {\bibfnamefont {Q.}~\bibnamefont {Quraishi}}, \bibinfo {author}
  {\bibfnamefont {S.}~\bibnamefont {Olmschenk}}, \bibinfo {author}
  {\bibfnamefont {W.}~\bibnamefont {Campbell}}, \bibinfo {author}
  {\bibfnamefont {J.}~\bibnamefont {Mizrahi}}, \bibinfo {author} {\bibfnamefont
  {C.}~\bibnamefont {Senko}}, \ and\ \bibinfo {author} {\bibfnamefont
  {C.}~\bibnamefont {Monroe}},\ }\href@noop {} {\bibfield  {journal} {\bibinfo
  {journal} {Phys. Rev. Lett.},\ }\textbf {\bibinfo {volume} {104}},\ \bibinfo
  {pages} {140501} (\bibinfo {year} {2010})}\BibitemShut {NoStop}%
\bibitem [{\citenamefont {Weiner}(1995)}]{WeinerPQE95}%
  \BibitemOpen
  \bibfield  {author} {\bibinfo {author} {\bibfnamefont {A.~M.}\ \bibnamefont
  {Weiner}},\ }\href@noop {} {\bibfield  {journal} {\bibinfo  {journal} {Prog.
  Quant. Electr.},\ }\textbf {\bibinfo {volume} {19}},\ \bibinfo {pages} {161}
  (\bibinfo {year} {1995})}\BibitemShut {NoStop}%
\bibitem [{\citenamefont {Olmschenk}\ \emph {et~al.}(2007)\citenamefont
  {Olmschenk}, \citenamefont {Younge}, \citenamefont {Moehring}, \citenamefont
  {Matsukevich}, \citenamefont {Maunz},\ and\ \citenamefont
  {Monroe}}]{OlmschenkPRA07}%
  \BibitemOpen
  \bibfield  {author} {\bibinfo {author} {\bibfnamefont {S.}~\bibnamefont
  {Olmschenk}}, \bibinfo {author} {\bibfnamefont {K.~C.}\ \bibnamefont
  {Younge}}, \bibinfo {author} {\bibfnamefont {D.~L.}\ \bibnamefont
  {Moehring}}, \bibinfo {author} {\bibfnamefont {D.~N.}\ \bibnamefont
  {Matsukevich}}, \bibinfo {author} {\bibfnamefont {P.}~\bibnamefont {Maunz}},
  \ and\ \bibinfo {author} {\bibfnamefont {C.}~\bibnamefont {Monroe}},\
  }\href@noop {} {\bibfield  {journal} {\bibinfo  {journal} {Phys. Rev. A},\
  }\textbf {\bibinfo {volume} {76}},\ \bibinfo {pages} {052314} (\bibinfo
  {year} {2007})}\BibitemShut {NoStop}%
\bibitem [{Note1()}]{Note1}%
  \BibitemOpen
  \bibinfo {note} {Coherent Paladin Compact 355-4000}\BibitemShut {NoStop}%
\bibitem [{Note2()}]{Note2}%
  \BibitemOpen
  \bibinfo {note} {Fast Pulse Technology model 5046SC}\BibitemShut {NoStop}%
\bibitem [{\citenamefont {Wineland}\ \emph {et~al.}(2003)\citenamefont
  {Wineland}, \citenamefont {Barrett}, \citenamefont {Britton}, \citenamefont
  {Chiaverini}, \citenamefont {De{M}arco}, \citenamefont {Itano}, \citenamefont
  {Jelenkovi\'{c}}, \citenamefont {Langer}, \citenamefont {Leibfried},
  \citenamefont {Meyer}, \citenamefont {Rosenband},\ and\ \citenamefont
  {Sch\"{a}tz}}]{WinelandPTRSLA03}%
  \BibitemOpen
  \bibfield  {author} {\bibinfo {author} {\bibfnamefont {D.~J.}\ \bibnamefont
  {Wineland}}, \bibinfo {author} {\bibfnamefont {M.}~\bibnamefont {Barrett}},
  \bibinfo {author} {\bibfnamefont {J.}~\bibnamefont {Britton}}, \bibinfo
  {author} {\bibfnamefont {J.}~\bibnamefont {Chiaverini}}, \bibinfo {author}
  {\bibfnamefont {B.}~\bibnamefont {De{M}arco}}, \bibinfo {author}
  {\bibfnamefont {W.~M.}\ \bibnamefont {Itano}}, \bibinfo {author}
  {\bibfnamefont {J.}~\bibnamefont {Jelenkovi\'{c}}}, \bibinfo {author}
  {\bibfnamefont {C.}~\bibnamefont {Langer}}, \bibinfo {author} {\bibfnamefont
  {D.}~\bibnamefont {Leibfried}}, \bibinfo {author} {\bibfnamefont
  {V.}~\bibnamefont {Meyer}}, \bibinfo {author} {\bibfnamefont
  {T.}~\bibnamefont {Rosenband}}, \ and\ \bibinfo {author} {\bibfnamefont
  {T.}~\bibnamefont {Sch\"{a}tz}},\ }\href@noop {} {\bibfield  {journal}
  {\bibinfo  {journal} {Phil. Trans. R. Soc. Lond. A},\ }\textbf {\bibinfo
  {volume} {361}},\ \bibinfo {pages} {1349} (\bibinfo {year}
  {2003})}\BibitemShut {NoStop}%
\bibitem [{\citenamefont {Ozeri}\ \emph {et~al.}(2007)\citenamefont {Ozeri},
  \citenamefont {Itano}, \citenamefont {Blakestad}, \citenamefont {Britton},
  \citenamefont {Chiaverini}, \citenamefont {Jost}, \citenamefont {Langer},
  \citenamefont {Leibfried}, \citenamefont {Reichle}, \citenamefont {Seidelin},
  \citenamefont {Wesenberg},\ and\ \citenamefont {Wineland}}]{OzeriPRA07}%
  \BibitemOpen
  \bibfield  {author} {\bibinfo {author} {\bibfnamefont {R.}~\bibnamefont
  {Ozeri}}, \bibinfo {author} {\bibfnamefont {W.~M.}\ \bibnamefont {Itano}},
  \bibinfo {author} {\bibfnamefont {R.~B.}\ \bibnamefont {Blakestad}}, \bibinfo
  {author} {\bibfnamefont {J.}~\bibnamefont {Britton}}, \bibinfo {author}
  {\bibfnamefont {J.}~\bibnamefont {Chiaverini}}, \bibinfo {author}
  {\bibfnamefont {J.~D.}\ \bibnamefont {Jost}}, \bibinfo {author}
  {\bibfnamefont {C.}~\bibnamefont {Langer}}, \bibinfo {author} {\bibfnamefont
  {D.}~\bibnamefont {Leibfried}}, \bibinfo {author} {\bibfnamefont
  {R.}~\bibnamefont {Reichle}}, \bibinfo {author} {\bibfnamefont
  {S.}~\bibnamefont {Seidelin}}, \bibinfo {author} {\bibfnamefont {J.~H.}\
  \bibnamefont {Wesenberg}}, \ and\ \bibinfo {author} {\bibfnamefont {D.~J.}\
  \bibnamefont {Wineland}},\ }\href@noop {} {\bibfield  {journal} {\bibinfo
  {journal} {Phys. Rev. A},\ }\textbf {\bibinfo {volume} {75}},\ \bibinfo
  {pages} {042329} (\bibinfo {year} {2007})}\BibitemShut {NoStop}%
\bibitem [{Note3()}]{Note3}%
  \BibitemOpen
  \bibinfo {note} {For Eq. \ref {OmegaRaman}-\ref {GammaSpon} we have ignored
  the small contributions from the $^3[3/2]^{\protect \mathrm {o}}_{3/2}$ state
  transition at 348 nm caused by the $^2\protect \mathsf {P}_{3/2}$ admixture
  in that state. We have also neglected the difference in transition dipole
  moment between the D1 and D2 lines. Both effects contribute less than 15\%
  corrections to the two-photon Rabi frequency, differential Stark shift, and
  spontaneous emission rate at 355 nm.}\BibitemShut {Stop}%
\bibitem [{Note4()}]{Note4}%
  \BibitemOpen
  \bibinfo {note} {Despite being slightly outside the fine-structure interval,
  a laser at 532 nm ($= 1064 \unhbox \voidb@x \hbox { nm}/2$) has a two-photon
  Rabi frequency for Ba$^+$ that is only slightly lower than Yb$^+$ at 355 nm
  with comparable pulse energy.}\BibitemShut {Stop}%
\bibitem [{\citenamefont {Rosen}\ and\ \citenamefont
  {Zener}(1932)}]{RosenPR32}%
  \BibitemOpen
  \bibfield  {author} {\bibinfo {author} {\bibfnamefont {N.}~\bibnamefont
  {Rosen}}\ and\ \bibinfo {author} {\bibfnamefont {C.}~\bibnamefont {Zener}},\
  }\href@noop {} {\bibfield  {journal} {\bibinfo  {journal} {Physical Review},\
  }\textbf {\bibinfo {volume} {40}},\ \bibinfo {pages} {502} (\bibinfo {year}
  {1932})}\BibitemShut {NoStop}%
\bibitem [{\citenamefont {Siegman}(1986)}]{Siegman}%
  \BibitemOpen
  \bibfield  {author} {\bibinfo {author} {\bibfnamefont {A.~E.}\ \bibnamefont
  {Siegman}},\ }\enquote {\bibinfo {title} {Lasers},}\ \ (\bibinfo  {publisher}
  {University Science Books},\ \bibinfo {year} {1986})\ Chap.~\bibinfo
  {chapter} {28}\BibitemShut {NoStop}%
\end{thebibliography}%
\end{document}